\documentclass[aps,prl,twocolumn,showpacs,10pt]{revtex4-1}

\usepackage{amssymb,amsmath,amstext}                
\usepackage{graphicx}                                               
\usepackage{epstopdf}                                               
\usepackage{color}                                                     
\usepackage{bm}                                                        
\usepackage{appendix}                                              
\usepackage[utf8]{inputenc}
\usepackage{bbold}
\usepackage{bbm}
\usepackage{ulem}
\usepackage[dvipsnames]{xcolor}

\usepackage{latexsym}
\usepackage[colorlinks=true,citecolor=blue,linkcolor=magenta]{hyperref}

\usepackage{filemod}

\def\be{\begin{equation}}
\def\ee{\end{equation}}
\def\bg{\begin{equation}\begin{gathered}}
\def\eg{\end{gathered}\end{equation}}

\def\up{\uparrow}
\def\dn{\downarrow}

\newcommand\Imrm{\mbox{Im\, }}

\begin{document}

\title{{\it Ab initio} study of time-dependent dynamics in strong field triple ionization}

\author{Jan H. Thiede and Bruno Eckhardt} 
\affiliation{Fachbereich Physik, Philipps-Universit\"at Marburg}

\author{Dmitry K. Efimov, Jakub S. Prauzner-Bechcicki and Jakub Zakrzewski}
\affiliation{Instytut Fizyki im. Mariana Smoluchowskiego, Uniwersytet Jagiello\'nski,  \L{}ojasiewicza 11, 30-348 Krak\'ow, Poland }
\email{jakub.zakrzewski@uj.edu.pl}

\begin{abstract}
An {\it ab initio} analysis of strong field three electron ionization in a restricted dimensionality model 
reveals the dynamics of the ionization process and the dominant channels for double (DI) and triple ionization (TI).
Simulations using wave functions that respect the Pauli principle show that the most likely channel is a 
sequence of single ionization (SI) and DI, while direct TI has a much lower probability.  
The dominant DI process has the highest probability for a singlet of up and down spin electrons. 
The results demonstrate the significance of the Pauli principle for the selection of dominant path
ways in ionization and possibly other many electron processes in strong fields.

\end{abstract}
\date{\today}

\maketitle

Ultrashort attosecond pulses enable {studies} of fundamental aspects of the interaction between
radiation and matter \cite{Krausz09,Calegari16,Ciappina17}. The generation of very high 
harmonics {in the process} is a key to the shaping of pulses and the realization of 
table-top sources of high frequency coherent radiation \cite{Popmintchev2010} that may 
compete with complex synchrotron or free electron laser  sources \cite{Bozek2009}.
High intensity pulses can also result in multiple ionization, typically
assisted by strong interactions between the escaping electrons \cite{Huillier83,Luk85,Larochelle98}.
Recently, the combination of both aspects, higher harmonic generation and multi-electron effects, 
has moved into focus \cite{Abanador18}. While there is an obvious need to study the 
strong-field multi-electron processes, 
a full {\it ab initio} computation remain limited to the case of 
two-electron atoms (helium) at high frequencies, as studied by Taylor's group \cite{Dundas99,Taylor03,Parker00,Emma11}, see also Refs. \cite{Feist08,Pazourek12},
despite the huge progress in theory and computer power.

In the absence of full simulations, approximate methods become important, and
semiclassical approaches 
\cite{Schafer93,Corkum93,Lewenstein94,Kubel16,Chen17,Milosevic17} 
as well as simulations in a variety of models have been explored. 
For double ionization (DI)  Rochester model, 
in which the motion of each electron is restricted to {the}  dimension {set} by the (linear) polarization of the laser field \cite{Grobe92}, has been studied.
The model was applied to illustrate, e.g., 
the mechanism of simultaneous ejection of two electrons at moderate intensity, 
and the transition to a sequential process for stronger fields 
(see e.g. \cite{Bauer97,Liu99,Lein00}). 
Despite its popularity, the model has its drawbacks: electrons moving 
in parallel directions repel each other
and this results in two-electron momentum distributions that disagree with observations. 
An {adiabatic} analysis allows one to  locate the saddles in the potential for the electrons in the
presence of an electric field, and such saddles act as transition states to efficient 
channels for ionization \cite{Eckhardt06}. That analysis led to the development of an
improved model in which electrons move along lines that pass through the saddles and 
are oblique to each other. The model takes electron
correlations into account and gives a plausible representation of the ionization process \cite{Prauzner07,Prauzner08}. 
A similar three-dimensional model is obtained by restricting the 
center-of-mass motion to the polarization axis; it captures similar aspects as the 
saddle picture (e.g. reproducing correctly the momenta distributions) 
\cite{Ruiz06,Staudte07,Chen10}, though with a larger number of degrees of freedom and at higher computational costs \cite{Efimov18}.

For the case of triple ionization (TI) studied here, several experimental results are available, especially for noble gases such as Kr, Ne or Xe \cite{Feuerstein00,Rudenko04,Palani05,Zrost06,Rudenko08,Ekanayake14}, 
but detailed theoretical studies are scarce,  because of the even larger number
of degrees of freedom.  Some isolated aspects 
have been described in {\it classical } studies \cite{sacha1,Ho06,Ho07,Guo08,Zhou10}, often within 
restricted dimensionality Rochester models.  
A notable quantum-mechanical effort \cite{Ruiz05,Ruiz06} considered TI of Li at high
frequencies corresponding to synchrotron radiation,
also within the Rochester model. Important progress has been made using different versions of multi-configuration Hartree-Fock 
time-dependent orbitals \cite{Anzaki17} (for a review see \cite{Ishikawa15}). 
This method, however,  depends on the number and an appropriate choice of initial orbitals included. Importantly, 
while it has been tested against quantum-mechanical results for the two-electron Rochester model, no such tests have been performed, 
as far as are aware, for three electron models since full quantum-mechanical analyses of 
the problem are still lacking.

The purpose of this work is to fill this gap, i.e. to provide a full 
{\it ab initio} quantum mechanical analysis of TI for {typical optical} frequencies within the reduced dimensionality scheme. For the Hamiltonian,  we consider models
motivated by the {adiabatic} analysis of the classical dynamics \cite{sacha1}, similar to  that described above for DI. 
As the electric field changes slowly compared to electron dynamics we consider the possible ionization channels as realized close to saddles of the instantaneous 
field value. The energetically lowest saddle corresponds to one with three electrons at the vertices of an equilateral triangle, in a plane perpendicular
to the field polarization axis \cite{sacha1}.
When the field amplitude is varied the saddles move along straight lines that point radially outwards from the core: 
in the restricted model, the motion of the electrons is confined to these lines.
As we will show, the analysis of the different sequential and simultaneous electron ejection processes 
provides a good understanding and explanation of the ionization yields.

\begin{figure*}
	\centering
	\includegraphics[width= 1.4\columnwidth]{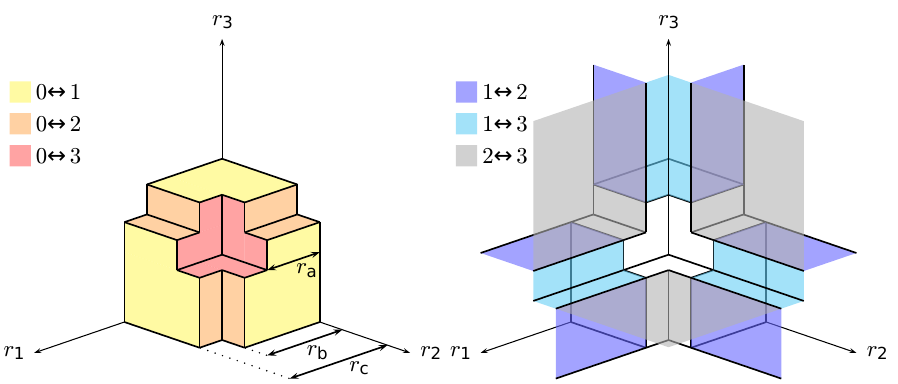}
	\caption{
		 Division of the position space for the calculation of the ion yields (only the first octant is shown). Region 0 (neutral atom) is the volume bounded by the yellow, orange and red planes. Region 1 (singly-ionized atom) is the union of the six volumes bounded by the yellow, cyan and blue planes. Region 2 (doubly-ionized atom) is the union of the twelve volumes bounded by the orange, blue and gray planes. Region 3 (triply-ionized atom) is the union of the eight volumes bounded by the red, cyan and gray planes. The missing boundary planes of regions 1-3 are given by the absorbing boundary (not shown).
	}\label{scheme}    
\end{figure*}  

\begin{figure*}
        \centering
        \includegraphics[width= 1.7\columnwidth]{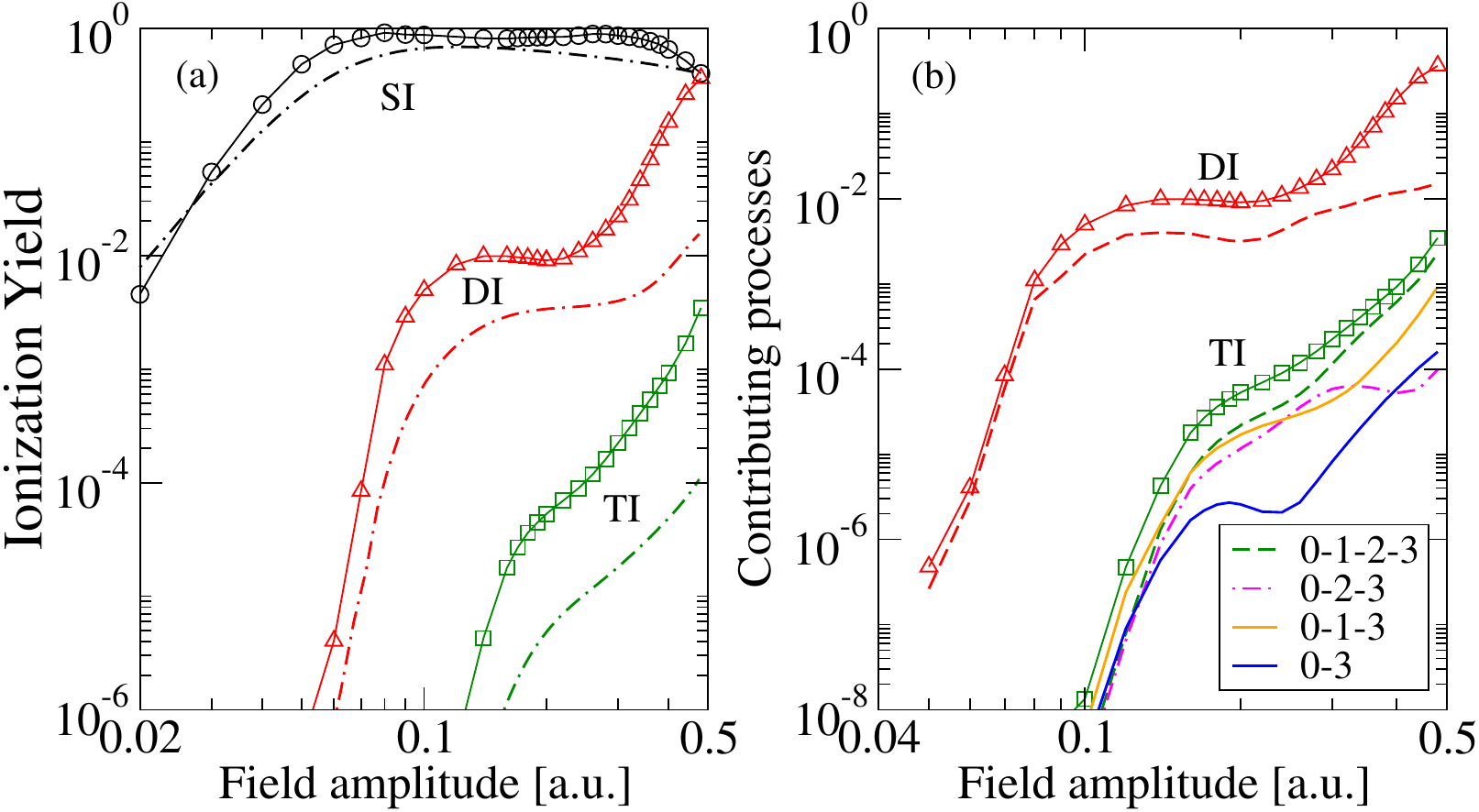}
        \caption{ Numerical ionization yields for a two-cycle pulse as a function of top electric field amplitude in atomic units. $F_0=0.1$~a.u. corresponds to $5.14\times 10^{10}$~V/m and laser intensity $I=3.5\times10^{14}$~W/cm$^2$. Left panel: total yields (probabilities) for single ionization (SI, black circles),
        double ionization (DI, red triangles) and triple ionization (TI, green squares). The continuous lines are from the model, 
        dash-dotted lines after integration over a Gaussian beam \eqref{aver}. SI saturates over a wide range of field strengths and 
        DI shows a pronounced knee structure, TI a weak one. 
        Right panel:  Different contributions to DI and TI, over a smaller range of field amplitudes. 
        The red dashed lines show the NSDI contribution to DI, which is overtaken by sequential processes for higher intensities. 
        The legend identifies four contributions to TI, with the sequential process being the dominant and direct TI the weakest contribution.
}\label{yield}
    \end{figure*}

The resulting Hamiltonian acting in an effective 3D space takes the form (in atomic units):
\be
H=\sum_{i=1}^3\frac{p_i^2}{2}+V(r_1,r_2,r_3)
\label{ham}
\ee
with 
\begin{eqnarray}
V(r_1,r_2,r_3)&=&-\sum_{i=1}^3\left(\frac{3}{\sqrt{r_i^2+\epsilon}} +\sqrt{\frac{2}{3}}F(t)r_i \right) \nonumber \\
&+&\sum_{i<j} \frac{1}{\sqrt{(r_i-r_j)^2+r_ir_j+\epsilon}}
\end{eqnarray}
where a parameter $\epsilon$ is responsible for smoothing of Coulomb singularity and, most importantly, allows us  to match the ionization potential of our model with those of the real atom under study. We consider the case of Ne for which several experimental studies are available (although for longer pulses)
\cite{Feuerstein00,Palani05,Zrost06,Rudenko08,Ekanayake14}.  More precisely, we consider a 3 active electron model of Ne, the remaining electrons are assumed to be spectators. The ground state energy of Ne  is $-4.63$a.u. \cite{Kramida16}, well approximated by the ground
state energy of $E_0=-4.619$ a.u. for $\epsilon=0.83$.

\begin{figure*}
      \centering
        \includegraphics[width= 1.8\columnwidth]{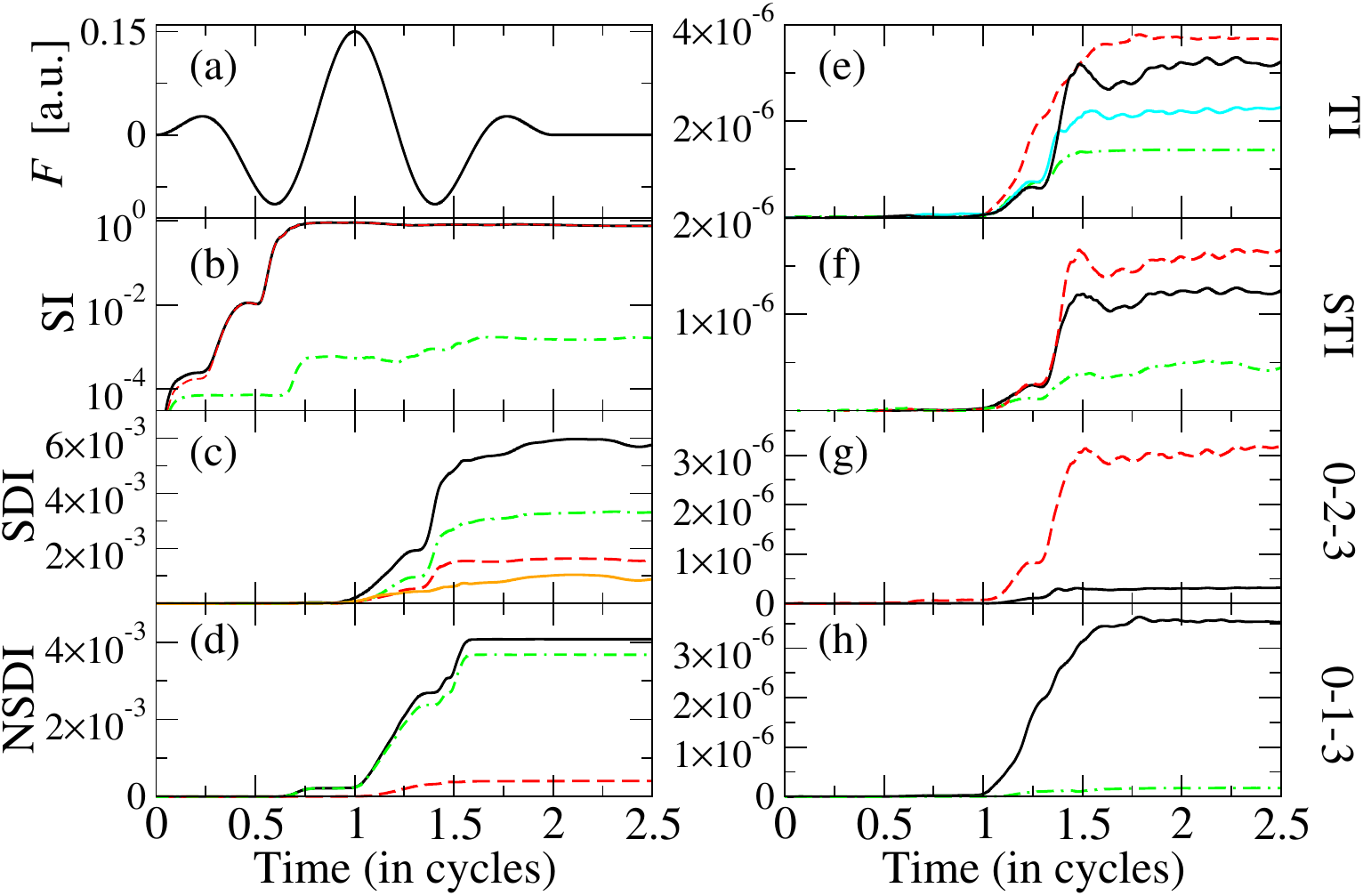}
        \caption{
Spin-resolved time-dependence of different ionization processes for 
$F_0=0.15$ and $\varphi=0$. The left column shows contributions to SI and DI, the right column to TI. 
Panel (a) shows the pulse shape. 
(b) SI (black line) is dominated by $U$ electron emission (red dashed), while ionization of 
$D$ electron (green dashed) has much smaller probability. 
(c) shows that sequential double ionization (SDI) (black line) is composed of the dominant $0$--$U$--$D$
channel (green dash-dotted), with only small contributions from the $0$--$U$--$U$ channel and minor
contributions from the path $0$--$D$--$U$ (orange)
in which $D$ electron is ejected first. 
For non-sequential double ionization (NSDI) (d) shows that the product  $DU$ (green dash-dotted) 
is strongly favored compared to $UU$ emission (red dashed). 
In the right column, (e) shows contributions to TI:  
sequential TI (STI) $0$--$1$--$2$--$3$ (cyan), $0$--$2$--$3$ channel (black), 
$0$--$1$--$3$ (red dashed line) 
and $0$--$3$ (green dashed line). 
Panel (f) resolves the spin contributions to STI, with black, red-dashed and green dot-dashed curves corresponding to $U$--$U$--$D$, $U$--$D$--$U$, 
and $D$--$U$--$U$ sequential emissions. 
(g) resolves the $0$--$2$--$3$ DI followed by single emission channel, 
and shows that the $DU$--$U$ path (red) is more prominent than the $UU$--$D$ sequence (black);
(h) for the $0$--$1$--$3$ channel the first SI is predominantly via $U$ electron followed by $UD$ pair (black) while
$D$--$UU$ (green dashed) is negligible.
}\label{temp}    \end{figure*}

The time-dependent Schr\"odinger equation  is solved on a spatial, equally spaced grid in three dimensions with Hamiltonian \eqref{ham} by a standard FFT 
(split-operator) technique in an efficiently parallelized way \cite{ThiedePhD}. The method is a straightforward generalization of our 
previous two-electron code \cite{Prauzner08} to three dimensions. However, accounting for the Pauli principle for
three electrons is more subtle than for two electrons. While for two electrons one may restrict the evolution to spaces that are
symmetric or antisymmetric under reflection of the position space wave functions \cite{Eckhardt08} this is not the case for three electrons. Writing a properly symmetrized wavefunction for three electrons as a product of spatial 
and spin parts in not possible. The correct three-electron wavefunction has to 
be constructed as a Slater determinant, which,
as shown  in Ref.\cite{Ruiz05}, reduces to
\begin{eqnarray}\label{wf}
\Psi_{\alpha\alpha\beta}(r_1,r_2,r_3,t) &\propto &\alpha(1)\alpha(2)\beta(3)\psi_{12}(r_1,r_2,r_3,t ) \nonumber \\ 
&+& \beta(1)\alpha(2)\alpha(3)\psi_{23}(r_1,r_2,r_3,t ) \\
&+& \alpha(1)\beta(2) \alpha(3)\psi_{13}(r_1,r_2,r_3,t ), \nonumber
\end{eqnarray}
where the single electron spin functions correspond to $\alpha(i)\equiv |\up\rangle_i$ and $\beta(i)\equiv |\dn\rangle_i$. To have a completely antisymmetric wavefunction 
$\psi_{ij}(r_1,r_2,r_3,t )$ is antisymmetric under exchange of $i$ and $j$. As pointed out in \cite{Ruiz06} all three components of 
$\Psi$ in the sum in Eq.~\eqref{wf} are orthogonal in spin space. Since the Hamiltonian \eqref{ham} is spin independent, all three terms 
in the sum evolve independently, so that it is enough to evolve a single one and to obtain the remaining two by appropriate change of indices.
Assuming the wavefunction to be antisymmetric in $r_1$ and $r_2$ we find the appropriate ground state in this symmetry class by an imaginary time propagation of TDSE, and this gives the ground state energy $E=-4.619$a.u. quoted above. 

We here consider ionization by an extremely short, 2-cycle pulse. While such short pulses are at the extreme limits of experimental availability, 
considerable progress towards their realization has been made
 \cite{Kim2013,Ossiander16,Kubel16,Calegari16,Ciappina17,Chen17}. The ionization process is then determined by few instances of time when the amplitude is large - that simplifies the detailed analysis of spin-dependent dynamics.  For such a short pulse it is imperative to construct the envelope in such a way that 
the vector potential, $A$, vanishes after the pulse has passed \cite{Eckhardt10}.
We therefore take $F(t) = -\partial A/\partial t$ with 
\begin{equation} \label{pulse}
A(t) = - \frac{F_0}{\omega_0}
\sin^2\left(\frac{\pi t}{T_p}\right) \sin(\omega_0 t+\varphi) 
\end{equation}
for $0<t<T_p$, where $\varphi$ defines the phase of the field under the pulse, 
$n_c$ the number of cycles and $T_p=2\pi n_c/\omega_0$ the pulse duration. 
{For the frequency, we take $\omega_0=0.06$, corresponding to a wavelength of $759$ nm.}

{For the determination of the yields, we split the configuration space into different sectors and 
compute the fluxes across the boundaries, in an extension of the procedures used in 
\cite{Dundas99,Ruiz05,Prauzner08}. 
The regions for the different states are composed of rectangular domains that are aligned with the coordinate axes, so 
that the boundaries between different regions consist of surfaces parallel to coordinate surfaces.
(compare Fig.~\ref{scheme}).
There is one region close to the nucleus, where all
electrons are bound. There are three regions extended along the coordinate axes where one of the
electrons is ionized and two are bound, 
another three regions where two electrons are free and one is bound
and finally a region far from the nucleus where all three electrons are free. 
For the distances defining the boundaries between the regions, we 
follow the idea of \cite{Dundas99} and take different defining distances that allow to distinguish
between simultaneous and step-wise DI and TI. 
We take
$r_c=12.5 {\rm a.u.}$ for SI,
$r_b=7{\rm a.u.}$ for DI, and $r_a=5{\rm a.u.}$ for TI -- Fig.~\ref{scheme}.
For instance, the region corresponding to a single charged ion 
has only one of the coordinates $r_i$ large ($r_i>r_c$). 
Similarly, we can identify regions corresponding to DI and TI as described in the caption of Fig.~\ref{scheme}.

{The yields are determined by integrating the fluxes between the regions. The fluxes 
are determined by integrating the probability currents, which are given by 
\begin{equation}
\textbf{j}(\textbf{r},t) = \Imrm (\psi^*(\textbf{r},t)\nabla\psi(\textbf{r},t))
\end{equation}
in length gauge or by 
\begin{equation}
\textbf{j}(\textbf{r},t) = \Imrm (\psi^*(\textbf{r},t)\nabla\psi(\textbf{r},t)) - \sqrt{2/3}| \psi(\textbf{r},t) |^2 A(t)
\end{equation}
in velocity gauge,  with vector potential $A(t)$. By Gauss's theorem, the fluxes determine the 
changes of the population in region $R \in \mathbb{R}^3$ according to
\begin{multline}
\frac{\partial}{\partial t} P_R (\textbf{r},t) = \frac{\partial}{\partial t} \iiint\limits_R |\psi(\textbf{r},t)|^2 \, d^3 \textbf{r} = \\ - \iiint\limits_R \nabla\cdot \textbf{j}(\textbf{r},t)  \, d^3 \textbf{r} = - \iint\limits_{\partial R} \textbf{j}(\textbf{r},t) \cdot d\mathbf{\sigma} \equiv f_R(t),
\label{flux}
\end{multline}
%
where $\partial R$ is the border of region $R$ and $d\mathbf{\sigma}$ 
is the corresponding surface element. 
We assume that the wavefunction decreases sufficiently rapidly as $r\rightarrow \infty$ 
so that all the above integrals 
converge for any region $R$. Correspondingly, the instantaneous value of the population in 
region $R$ is given by
\begin{equation}
P_R (\textbf{r},t) = P_R (\textbf{r},0) - \int_0^t f_R(t')\, dt'.
\label{prob}
\end{equation}
We have checked that changes  of the defining distances affect the  ionization probabilities quantitatively only, leaving the qualitative picture, which is our aim with the reduced dimensionality model, 
unaffected. By considering which electron travels outside  the bound state region we can identify which spin channels are most vulnerable to ionisation - recall that the antisymmetric configuration space wave-function corresponds to a majority spin pointing up.}

Let us first consider the yields obtained for a  pulse (\ref{pulse}) for different values of the field peak amplitude $F_0$, as shown in Fig.~\ref{yield} and $\varphi=0$.
One observes a  fast saturation 
of  SI  that reveals upon close inspection of the data, a shallow maximum around $F_0=0.2$ followed by a decay for larger amplitudes when DI and then TI become important.  
 From the contributions to the DI yield,  we can determine the ratio of $0$--$1$--$2$
(sequential DI) to $0$--$2$ (NSDI). 
Similarly, for TI we may define sequences like $0$--$2$--$3$ or
 $0$--$1$--$3$,  corresponding to combinations of SI and DI in different orders,  or $0$--$3$, 
 a simultaneous TI process. 
 Note that the flux method does not allow us to distinguish a  sequential $0$--$1$--$2$--$3$ 
 process from a nonsequential $0$--$2$--$3$ scenario, since the integrated flux across the 
 $2$--$3$ border 
determines the $0$--$2$--$3$ process which contains in the 
$0$--$2$ part both the sequential $0$--$1$--$2$ and a direct $0$--$2$ path. 
However, we know the effectiveness of $0$--$1$--$2$ 
versus $0$--$2$ channel from the corresponding fluxes,
and 
assuming that the same ratio holds for three electron processes , 
we may deduce approximate values for the corresponding yields 
in TI. 
The results for the yields and the different contributions for a two-cycle
pulse are shown in Fig.~\ref{yield}.

In the experiment atoms are illuminated 
by a Gaussian laser beam. In the computations, this can be accounted for by averaging the yields over the laser beam intensity profile. 
As shown in \cite{Strohaber15},  the  averaged ionization yields
$S(I_0)$ may simply be obtained as
\be\label{aver}
 S(I_0)\propto \int_0^{I_0} dI P(I)/I 
\ee
where $I_0\propto F_0^2$ is the peak intensity at the focal point and $P(I)$ is a fraction obtained numerically for a given peak intensity $I$.
The results of such an averaging are shown as 
dash-dotted lines in  the left panel of Fig.~\ref{yield}. Note that the knee structure, indicating the transition
from the non-sequential processes to the sequential ones, becomes significantly 
smoothed out and the resulting average yields resemble qualitatively
the ones observed in experiments for Ne  \cite{Feuerstein00,Palani05,Zrost06,Rudenko08,Ekanayake14}.

The dominant feature in Fig.~\ref{yield} is a deep knee structure for a DI, mostly due 
to NSDI (as indicated by red curves).
This happens in the same interval of field amplitudes as the saturation, together with 
a small drop of the single ionization yield. For stronger fields, the fraction of NSDI becomes less significant in the total DI yield, 
and we recover  
the sequential path familiar from earlier studies. For even stronger intensities, TI sets in with less pronounced saddles. 
Note that direct TI is the least probable scenario, with DI (either sequential or NSDI) followed by a SI process
being the most effective process. 

The access to the time-dependent fluxes across the different borders 
also provides information
about the spin-polarization of the outgoing electrons.
Recall that our three-electron initial wavefunction is composed of two spin-up electrons (here denote by $U$) and one 
spin-down electron (denoted by $D$). The wavefunction is antisymmetric with respect to the exchange of $U$ electrons, and
symmetric with respect to an exchange between $U$ and $D$ electrons. The fluxes allow us to address the question 
whether it is more probable to eject first a $U$ or a $D$ electron. Intuition suggests that if one of the $U$'s and $D$ form a singlet, 
the remaining $U$ electron is easier to ionize. And indeed, the SI yield for the $D$ electron is negligible! (compare Fig.~\ref{temp}). 
Since our approach gives us a direct access to time-dependent fluxes by defining appropriate ionization processes, we can in a 
similar way analyze DI and TI events. In particular, such an analysis points towards $DU$ emission as a dominant channel for NSDI, with
simultaneous emission of two $U$ electrons being much less probable. Similarly, we may identify the 
dominant channels for TI. After splitting the $0$--$2$--$3$ channel into the sequential 
$0$--$1$--$2$--$3$ and NSDI followed by single electron emission, 
the leading channel becomes $0$--$1$--$3$ for low field amplitude. In such a case of SI followed by simultaneous ejection of the remaining two electrons,  
the first stage is almost surely performed by the $U$ electron. On the other hand, 
the often neglected $0$--$2$--$3$ channel 
\cite{Feuerstein00} may be the leading TI channel for intermediate field values. 
All the possible channels are described in Fig.~\ref{temp} caption. 

\begin{figure}
      \centering
        \includegraphics[width= 0.95\columnwidth]{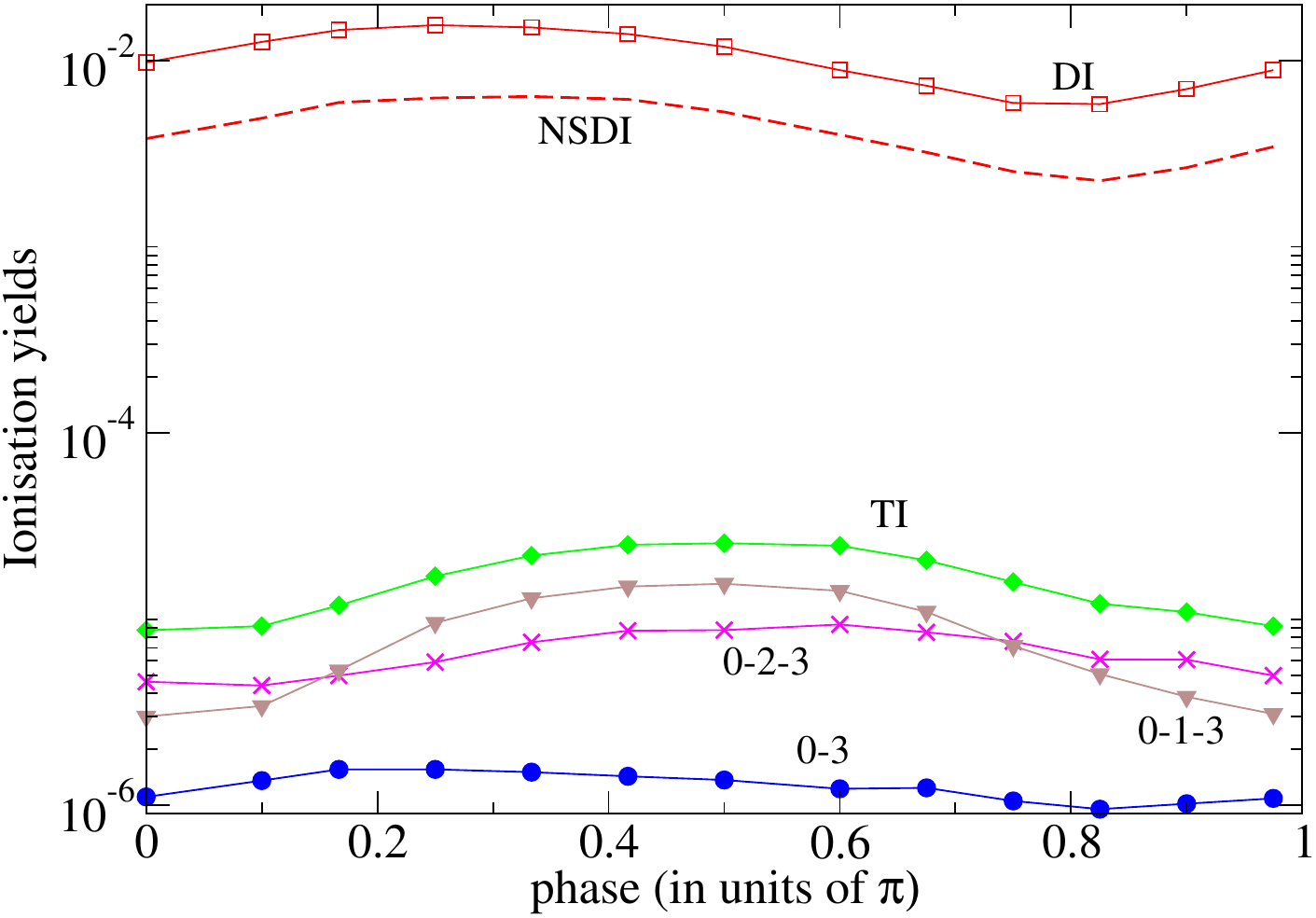}
        \caption{ The dependence of different ionization yields (indicated in the Figure) on the carrier envelope phase $\varphi$ in Eq.~\eqref{pulse}.  
        The data correspond to $F_0=0.15$. Observe that the importance of different paths for triple ionization may depend on  $\varphi$.
}\label{figphase} 
    \end{figure}
    
For short 2-cycle pulse used in calculations the shape and maximal amplitude (for a given $F_0$) depend 
on the carrier envelope phase (CEP) $\varphi$, see \eqref{pulse}. The effects on the yields are shown in
Fig.~\ref{figphase}. One observes that CEP values for the most effective DI and TI are different. Moreover,
the efficiency of different TI channels depends on CEP, e.g. the efficiency of $0-1-3$ and $0-2-3$
TI channels may be reversed (we here do not separate the $0-2-3$ process further for simplicity). 
Regardless of the CEP value the direct $0-3$ ionization channel is the least effective.
On the other hand, the main feature, i.e., that $U$ (majority population) electrons ionize first, does not depend on details 
of the pulse. Similarly, in non-sequential processes, it is a ``singlet'' pair $UD$ which is more likely to be ejected than a $UU$ combination.

The present study paves the way towards a detailed analysis of dynamics of three active electron dynamics for  Li as well as other noble gases and for longer pulses. While we have concentrated on the ionization yield and the dynamics of the process, work is in progress concerning the high order harmonic generation and ion momenta distribution analysis.

\acknowledgments
The support  of PL-Grid Infrastructure essential for obtaining the numerical results presented is acknowledged. We acknowledge support of the National Science Centre, Poland via project No.2016/20/W/ST4/00314 (DE, JPB and JZ). Support by EU FET-PRO QUIC is also acknowledged.

\providecommand{\noopsort}[1]{}\providecommand{\singleletter}[1]{#1}%

\end{document}